\documentclass[10pt,a4paper]{article}

\usepackage[T1]{fontenc }
\usepackage{amsmath}
\usepackage{amscd}
\usepackage{amsfonts}
\usepackage{amssymb}
\usepackage{bm}
\usepackage{graphicx}
\usepackage{geometry}

\newtheorem{exmp}{Example}
\newtheorem{prop}{Proposition}

\def\1{{\mathchoice{\rm 1\mskip-4mu l}{\rm 1\mskip-4mu l}%
{\rm 1\mskip-4.5mu l}{\rm 1\mskip-5mu l}}}
\def\la{\langle}
\def\ra{\rangle}
\def\ket#1{|#1\ra}
\def\bra#1{\la #1|}

\def\mathbm#1{\mbox{\boldmath$#1$}}
\def\pfrac#1#2{\frac{\partial #1}{\partial #2}}

\title{Local numerical range for a class of $2\otimes
d$ hermitian operators}
\author{J. Jurkowski, A. Rutkowski, D. Chru\'sci\'nski\\[1ex]
Institute of Physics, Nicolaus Copernicus University \\
Grudziadzka 5, 87--100 Toru\'n, Poland}
\date{}
\begin{document}
\maketitle
\begin{abstract}
A local numerical range is analyzed for a family of circulant
observables and states of composite $2 \otimes d$ systems. It is
shown that for any $2\otimes d$ circulant operator $\cal O$ there
exists a basis giving rise to the matrix representation with real
non-negative off-diagonal elements. In this basis the problem of
finding extremum of $\cal O$ on product vectors $\ket{x}\otimes
\ket{y} \in \mathbb{C}^2\otimes \mathbb{C}^d$  reduces to the
corresponding problem in $\mathbb{R}^2\otimes \mathbb{R}^d$. The
final analytical result for $d=2$ is presented.

\end{abstract}

\section{Introduction}

For any linear operator $\mathcal{O}$ acting in the Hilbert space
$\mathcal{H}$ one defines its {\em numerical range} \cite{HJ}
\begin{equation}\label{}
    {\rm NR}(\mathcal{O}) := \{\ \bra{\psi}   \mathcal{O} \ket{\psi} \ | \
    \psi \in \mathcal{H}\ , \
    ||\psi||=1\ \} \ .
\end{equation}
Clearly, ${\rm NR}(\mathcal{O})$ defines a subset of the complex
plane. Now, if $\mathcal{O}$ is hermitian then ${\rm
NR}(\mathcal{O}) = [ \lambda_{\rm min},\lambda_{\rm max}]$, where
$\lambda_{\rm min}$ and $\lambda_{\rm max}$ denote the minimal and
maximal eigenvalue of $\mathcal{O}$. Recently, more specific
characterization of the hermitian operator called {\it restricted
numerical range\/}  has been introduced in order to describe the
interval of expectation values for some specific sets of vectors in
$\mathcal{H}$ \cite{Z-1}. In particular, if $\mathcal{H} =
\mathcal{H}_1 \otimes \mathcal{H}_2$ one introduces the notion of
{\it local (product) numerical range\/} \cite{Z-2}
\begin{equation}
    {\rm LNR}({\cal O})\;=\;\{\la x\otimes y|{\cal O}|x\otimes y\ra\,:\,
    ||x||=||y||=1\,,\;
    \ket{x}\in\mathcal{H}_1\,,\;\ket{y}\in\mathcal{H}_2 \}\,.
\end{equation}
It is clear that if $\mathcal{O}$ is hermitian then
\[  {\rm LNR}({\cal O}) = [\gamma_{\rm
min},\gamma_{\rm max}]\ \subseteq\  {\rm NR}({\cal O}) =
[\lambda_{\rm min},\lambda_{\rm max}]\ . \] It turns out that the
notions of various restricted numerical ranges are useful in many
branches of quantum information theory (see \cite{Z-1,Schulte,Dirr}
for details). For example any entanglement witness $W$ can be
written in the following form \cite{Toth,Toth-2,Hor09,vogel}
\[
W=\chi\1- {\cal O}\,,
\]
for some hermitian operator ${\cal O}$ and a positive number $\chi$.
Now, the necessary condition for $W$ to be an entanglement witness
is $\chi > \gamma_{\rm max}$.  In practice, it is very hard  to
determine LNR for a given hermitian operator. In this paper we limit
ourselves to the case when ${\cal O}$ acting on
$\mathbb{C}^2\otimes\mathbb{C}^d$ belongs to a class of circulant
operators \cite{darek} (see also \cite{Art,Multi}).

The paper is organized as follows. Sect.~2 is devoted to some basic
definitions and properties of circulant bipartite operators. In
Sect.~3, we emphasize that it is always possible to bring a matrix
representing the circulant operator to the so-called {\it real\/}
form using a local unitary transformation. In Sect.~4 we show how to
carry out calculations of the local numerical range for circulant
operators. The final analytical result for $d=2$ is presented in
Sect.~5 together with some instructive examples.

\section{Circulant operators in $\mathbb{C}^2\otimes\mathbb{C}^d$}

Let $\mathcal{H} = \mathbb{C}^2\otimes\mathbb{C}^d$ and let
$\{\ket{g_i}\otimes\ket{f_k}\}$ ($i=1,2$, $k=1,\ldots,d$) be  an
orthonormal product basis in $\mathcal{H}$. One defines the family
of $2$-dimensional subspaces $\Sigma_k$ in $\mathcal{H}$:
\begin{eqnarray*}
\Sigma_1 &=& {\rm span}\Big\{\ket{g_1}\otimes \ket{f_1},\ket{g_2}\otimes \ket{f_2}\Big\}\,, \\
\Sigma_2 &=& {\rm span}\Big\{\ket{g_1}\otimes
\ket{f_2},\ket{g_2}\otimes \ket{f_3}\Big\}
\\
 & \vdots & \\
\Sigma_d &=& {\rm span}\Big\{\ket{g_1}\otimes
\ket{f_d},\ket{g_2}\otimes \ket{f_1} \Big\}\,.
\end{eqnarray*}
It is clear that $\Sigma_k$ give rise to the direct sum
decomposition \cite{darek,darek2}
\begin{equation}\label{circ-decomp}
    \mathbb{C}^2\otimes\mathbb{C}^d\;=\;\bigoplus_{k=1}^d\Sigma_k\,
    .
\end{equation}
We shall call (\ref{circ-decomp}) a {\em circulant decomposition}.
Now, we call a linear operator $\mathcal{O} \in
\mathcal{B}(\mathcal{H})$ to be {\em circulant operator} withe
respect to a circulant decomposition (\ref{circ-decomp}) iff
\begin{equation}\label{}
\mathcal{O} = \mathcal{O}_1 \oplus \ldots \oplus \mathcal{O}_d\ ,
\end{equation}
where $\mathcal{O}_k$ is supported on $\Sigma_k$, that is,
\begin{equation}\label{}
    \mathcal{O}_k = \sum_{i,j=1}^2a^{(k)}_{ij}
\ket{g_i}\bra{g_j}\otimes \ket{f_{i+k}}\bra{f_{j+k}}\,, \ \ \ \ {\rm
mod} \ d\ ,
\end{equation}
and $||a^{(k)}_{ij}||$ is a  $2 \times 2$ complex matrix.  In
particular, for $d=2$ and $d=3$ we obtain the following matrix
representations of the circulant operators (in the basis
$\ket{g_i}\otimes\ket{f_k})$
\begin{equation}
%    M^{2\otimes2}=
\left(\begin{array}{cccc}
    a^{(2)}_{11} &\cdot &\cdot &a^{(2)}_{12} \\
    \cdot & a^{(1)}_{11}& a^{(1)}_{12} & \cdot \\
        \cdot & a^{(1)}_{21}& a^{(1)}_{22} & \cdot \\
        a^{(2)}_{21} &\cdot &\cdot & a^{(2)}_{22}
        \end{array}\right),\quad
%        M^{2\otimes3}=
\left(\begin{array}{cccccc}
    a^{(3)}_{11} &\cdot &\cdot &\cdot &a^{(3)}_{12} &\cdot \\
    \cdot & a^{(1)}_{11}&\cdot&\cdot&\cdot& a^{(1)}_{12} \\
        \cdot &\cdot& a^{(2)}_{11}& a^{(2)}_{12} & \cdot &\cdot \\
        \cdot&\cdot&a^{(2)}_{21}&a^{(2)}_{22}&\cdot&\cdot \\
        a^{(3)}_{21} &\cdot &\cdot&\cdot & a^{(3)}_{22} &\cdot \\
        \cdot &a^{(1)}_{21}&\cdot&\cdot&\cdot&a^{(1)}_{22}
        \end{array}\right)\ ,
\end{equation}
where to make the picture more transparent we replaced all zeros by
dots. Interestingly for $d=2$ the circulant matrix displays
characteristic X-shape. Such 2-qubit states  have been recently
investigated in \cite{Rau,Fan10,Maz,Bogna,YWei}. In the following we
limit ourselves to circulant states and observables, i.e.~hermitian
circulant matrices, only. Let us introduce a more convenient
notation and denote by
\[  w_{ik}=a_{ii}^{(k-i)}\ , \ \ \  {\rm mod}\ d \ ,
\]
for $i=1,2$, $k=1,\ldots,d$, and \[
  a_{12}^{(k+2)}=u_k e^{i\alpha_k}  \ , \]
where $u_k = |a_{12}^{(k+2)}| \geq 0$, and $\alpha_k\in(-\pi,\pi]$.
As a consequence, the general circulant observable reads
\begin{equation}\label{circ-obs}
    {\cal O}=\sum_{i=1}^2\sum_{k=1}^dw_{ik}\ket{g_i}\bra{g_i}\otimes\ket{f_k}\bra{f_k}+
    \Big(\sum_{k=1}^du_ke^{i\alpha_k}\ket{g_1}\bra{g_2}\otimes\ket{f_k}\bra{f_{k+1}}+{\rm
    h.c.}\Big) \ ,
\end{equation}
where as usual h.c. stands for hermitian conjugation.

\section{Real representation of circulant operators}

Let $\mathcal{O}$ be an hermitian circulant operator living in
$\mathbb{C}^2 \otimes \mathbb{C}^d$. One has the following

\begin{prop}
There exists an orthonormal product basis
$\{\ket{g_i'}\otimes\ket{f_k'}\}$ such that

\begin{enumerate}

\item $\mathcal{O}$ is circulant with respect to the circulant
decomposition constructed out of $\{\ket{g_i'}\otimes\ket{f_k'}\}$,

\item matrix elements of $\mathcal{O}$ with respect to
$\{\ket{g_i'}\otimes\ket{f_k'}\}$ satisfy:

\begin{equation}\label{}
    w_{ik}' = w_{ik} \ , \ \ \ {a_{12}^{(k+2)}}'= |a_{12}^{(k+2)}|= u_k
    \ .
\end{equation}

\end{enumerate}

\end{prop}

\noindent {\it Proof}. Let $\ket{g_i'} = U_1 \ket{g_i}$ and
$\ket{f_k'} = U_2 \ket{f_k}$, where $U_1$ and $U_2$ are unitary
operators with the following matrix representations in the original
basis $\ket{g_i}$ and $\ket{f_k}$:
\begin{equation}\label{}
    U_1 = D[1,e^{i\mu_1}]\ , \ \ \ U_2 =
    D[1,e^{i\mu_2},\ldots,e^{i\mu_d}]\ ,
\end{equation}
where $D[a_1,\ldots,a_k]$ denotes diagonal $k\times k$ matrix with
diagonal entries $a_1,\ldots,a_k$. One has
\begin{eqnarray}  \label{O-new}
{\cal O} =
\sum_{i=1}^2\sum_{k=1}^dw_{ik}\ket{g_i'}\bra{g_i'}\otimes\ket{f_k'}\bra{f_k'}+
    \Big(\sum_{k=1}^du_ke^{i\vartheta_k}\ket{g_1'}\bra{g_2'}\otimes\ket{f_k'}\bra{f_{k+1}'}+{\rm h.c.}\Big),
\end{eqnarray}
where the phases $\vartheta_k$ satisfying the following relations
(mod$(2\pi)$)
\begin{eqnarray} \label{fazy}
    \vartheta_1 &=& \alpha_1-\mu_1-\mu_2\,, \nonumber \\
    \vartheta_k &=& \alpha_k-\mu_1+\mu_k-\mu_{k+1}\,,\quad k=2,\ldots,d-1 \\
    \vartheta_d &=& \alpha_d-\mu_1+\mu_d\,.  \nonumber
\end{eqnarray}
Formula (\ref{O-new}) proves that $\mathcal{O}$ is circulant with
respect the circulant decomposition constructed out of
$\{\ket{g_i'}\otimes\ket{f_k'}\}$. Now, we show that one can remove
all the phases $\vartheta_k$ by the appropriate choice of $\mu_k$.
Note, that (\ref{fazy}) may be rewritten as a matrix equation
$\mathbm{\alpha}-\mathbm{\vartheta}=\mathbm{W}\mathbm{\mu}$, where
the matrix $\mathbf{W}$ is defined by
\begin{eqnarray}\label{W}
    W_{k1} & = &1\ , \nonumber \\
    W_{kk} &=& - W_{k,k+1} \ , \ \ \ k>1\ ,
\end{eqnarray}
and the remaining elements vanish. Note that  taking $d$-vector
$\mathbm{\mu}=(\mu_1,\ldots,\mu_d)$ which satisfies the matrix
equation
\begin{equation}\label{mW}
    \mathbm{\alpha}=\mathbm{W}\mathbm{\mu}\,,
\end{equation}
one finds $\mathbm{\vartheta}=0$. It can be done due to the fact
that ${\rm det}\,\mathbm{W}=d(-1)^{d+1}\neq0$ which ends the proof.
\hfill $\Box$

We will call the corresponding matrix representation of
$\mathcal{O}$ with respect to $\{\ket{g_i'}\otimes\ket{f_k'}\}$ {\it
real representation}.

\section{Local Numerical Range for a Circulant Operator}

Let $\mathcal{O}$ be an hermitian circulant operator with respect to
a fixed basis $\ket{g_i} \otimes \ket{f_k}$ in $\mathbb{C}^2 \otimes
\mathbb{C}^d$, and let us define
\begin{equation}\label{F}
F(x,y)= \frac{\la x\otimes y|{\cal O}|x\otimes y\ra}{\la x\otimes
y|x\otimes y\ra}  \ .
\end{equation}
Now to provide LNM$(\mathcal{O})$ one has to find $\gamma_{\rm min}
= {\rm inf}\,F(x,y)$ and $\gamma_{\rm max} = {\rm sup}\,F(x,y)$. Let
\begin{equation}\label{}
\gamma_{\rm min} = F(x^-,y^-)\ ,\ \ \ \  \gamma_{\rm max} =
F(x^+,y^+)\ .
\end{equation}
One has the following

\begin{prop} The corresponding vectors $\ket{x^\pm} \in \mathbb{C}^2$ and $\ket{y^\pm}
\in \mathbb{C}^d$ have the following components with respect to
basis $\ket{g_i'}$ and $\ket{f_k'}$ provided in Proposition 1
\begin{equation}\label{}
    \ket{x^{\pm}} = (x^\pm_1,x^\pm_2) \ , \ \ \
    \ket{y^\pm}=(y^\pm_1,\ldots,y^\pm_d)\ ,
\end{equation}
where
\begin{equation}\label{}
    x_i^\pm \geq 0 \ , \ \ \ y^\pm_k \geq 0\ .
\end{equation}

\end{prop}

\noindent {\it Proof}. Consider e.g. $\gamma_{\rm min}$ and to
simplify notation let us write simply $\ket{x}$ and $\ket{y}$
instead of $\ket{x^-}$ and $\ket{y^-}$, respectively. Moreover, let
us introduce the following parametrization of vectors $\ket{x} \in
\mathbb{C}^2$ and $\ket{y}\in \mathbb{C}^d$ in the original basis
$\ket{g_i} \otimes \ket{f_k}$:
\begin{equation}\label{vec-max}
\ket{x}=(x_1,x_2e^{i\beta_1})\,,\quad
\ket{y}=(y_1,y_2e^{i\beta_2},\ldots,y_de^{i\beta_d})\,,\quad x_1,x_2\geq 0\,,\quad y_1,\ldots,y_d\geq 0\,.
\end{equation}
Using (\ref{circ-obs}) one obtains
\begin{eqnarray}\label{forma}
\la x\otimes y|{\cal O}|x\otimes y\ra &=&
\sum_{i=1}^2\sum_{k=1}^dw_{ik}x_i^2y_k^2+2x_1x_2\sum_{k=1}^d
    y_ky_{k+1}u_k\cos\varphi_k\,,
\end{eqnarray}
where
\begin{eqnarray}
    \varphi_1 &=& \alpha_1+\beta_1+\beta_2\,,\nonumber\\
    \varphi_k &=& \alpha_k+\beta_1-\beta_k+\beta_{k+1}\,,\quad k=2,\ldots,d-1 \\
    \varphi_d &=& \alpha_d+\beta_1-\beta_d\,. \nonumber
\end{eqnarray}
The extremalization procedure leads to the set of equations for real
positive variables $x_1, x_2$, $y_1,\ldots,y_d$ and for the phases
$\beta_1,\ldots, \beta_d$ (see Appendix for details). In particular,
phases $\beta_k$ can be easily obtained in the generic case, i.e.,
for $x_i\neq 0$,  and $y_k\neq 0$, as shown in (\ref{sol-beta}).
Using simple algebra (see the Appendix) one finds
\begin{equation}\label{}
\beta_k=-\mu_k\ , \ \ \ \ k=1,\ldots,d,
\end{equation}
where $\mu_k$ are solutions of (\ref{mW}). Hence in the new basis
$\ket{g_i'} \otimes \ket{f_k'}$ the phases $\beta_k$ are completely
removed and the components of $\ket{x}$ and $\ket{y}$ are
non-negative. \hfill $\Box$

Hence, essentially  LNR$(\mathcal{O})$  calculations can be  done in
 $\mathbb{R}^2\otimes\mathbb{R}^d$ instead of $\mathbb{C}^2\otimes\mathbb{C}^d$ . Unfortunately, solving the set
of $d+2$ polynomial equations (\ref{poly-x}), (\ref{poly-y}) is in
general very hard. Keeping in mind that in the basis
$\{\ket{g_i'}\otimes \ket{f_k'}\}$ all $\varphi_k=0$, we can rewrite
(\ref{poly-x}) as
\begin{equation}\label{eq-x}
\left\{\begin{array}{rcl}
(A_1(y)-\lambda_1)x_1+B(y)x_2 &=& 0\,,\\
B(y)x_1 +(A_2(y)-\lambda_1)x_2 &=& 0\,,
\end{array}\right.
\end{equation}
with
\[
A_\ell(y)=\sum_{k=1}^dw_{\ell k}y_k^2\,,\qquad B(y)=\sum_{k=1}^du_ky_ky_{k+1}\,.
\]
Now, we obtain the nonzero solution for $x_1$, $x_2$ from a linear set of equations
(\ref{eq-x}) if
\[
\lambda_1^{\pm}=\frac12\Big(A_1(y)+A_2(y)\pm\sqrt{(A_1(y)-A_2(y))^2+4B(y)^2}\Big)\,.
\]
Let us write this solution as
\begin{equation}\label{sol-x}
 x_1=\frac1{\sqrt{1+C_{\pm}^2}}\,,\qquad x_2=C_{\pm}x_1=
\frac{C_{\pm}}{\sqrt{1+C_{\pm}^2}},
 \end{equation}
where
\[
C_{\pm}=\frac{\lambda_1^{\pm}-A_1(y)}{B(y)}
\]
and the normalization of $\ket{x}$ has been taken into account.
Putting (\ref{sol-x}) into (\ref{poly-y}) we arrive at the following
set of $d$ nonlinear equations for $y_1,\ldots,y_d$:
\begin{equation}\label{set-y}
\Big[\frac{1}{C_{\pm}}w_{1k}+C_{\pm}w_{2k}-\lambda_2\frac{1+C_{\pm}^2}{C_{\pm}}\Big]y_k
+u_{k-1}y_{k-1}+u_ky_{k+1}=0\,,\quad k=1,\ldots,d\,.
\end{equation}
Clearly, in general the solution of (\ref{set-y}) is not feasible.
Note however that when $A_1(y)=A_2(y)$, i.e.~$w_{1k}=w_{2k}$ for
$k=1,\ldots,d$, one gets $C_{\pm}=\pm 1$ and the set of equations
(\ref{set-y}) becomes linear.

\begin{exmp}\rm Let us consider circulant hermitian operator $\mathcal{O}$ in $\mathbb{C}^2 \otimes \mathbb{C}^2$
represented in the standard computational basis by the following
real matrix
\begin{equation}
    M_{\cal O}\;=\;\left(\begin{array}{cccc}
2 & 0 & 0 & 1 \\
0 & 1 & -1 & 0 \\
0 & -1 & 1 & 0 \\
1 & 0 & 0 & 2 \\
\end{array}\right).
\end{equation}
The spectrum of $M_{\cal O}$ is $\{0,1,2,3\}$. As a consequence,
NR$({\cal O})=[0,3]$, whereas, as we shall see, LNR$({\cal
O})=[0.5,2.5]$. Moreover, the upper bound $\gamma_{\rm max}$ is
achieved at complex vectors $\ket{x}=\frac{1}{\sqrt{2}}(1,i)$ and
$\ket{y}=\frac{1}{\sqrt{2}}(1,-i)$ and when calculating expectation
values on normalized vectors from $\mathbb{R}^2\otimes\mathbb{R}^2$
we do not go beyond 2.

In order to proof that the upper bound of LNR$({\cal O})$ is indeed
2.5, let us bring the observable ${\cal O}$ into the real form by a
local unitary transformation (which does not change the ranges but
does change the extremal vectors),
\begin{equation}
    U_1 = D[1,-i] \ , \ \ \ \ U_2=D[1,i]\ .
\end{equation}
giving rise to
\begin{equation}
    M_{\cal O}'\;=\; U_1 \otimes U_2 M_O U_1^\dagger
\otimes U_2^\dagger = \left(\begin{array}{cccc}
2 & 0 & 0 & 1 \\
0 & 1 & 1 & 0 \\
0 & 1 & 1 & 0 \\
1 & 0 & 0 & 2 \\
\end{array}\right).
\end{equation}
Now, it is easy to show that for $\ket{x}=(x_1,x_2)\in\mathbb{C}^2$
and $\ket{y}=(y_1,y_2)\in\mathbb{C}^2$ we get
\begin{eqnarray}
    \la x\otimes y|M_{\cal O}'|x\otimes y\ra&=&2(|x_1|^2|y_1|^2+|x_2|^2|y_2|^2)+|x_1|^2|y_2|^2+|x_2|^2|y_1|^2+
    4{\rm Re}(x_1x_2^*){\rm Re}(y_1y_2^*)\nonumber\\
    &\leq& 2(|x_1|^2|y_1|^2+|x_2|^2|y_2|^2)+|x_1|^2|y_2|^2+|x_2|^2|y_1|^2+1\label{eq}
    \end{eqnarray}
due to ${\rm Re}(x_1x_2^*)\leq 1/2$ which follows from the
normalization condition $|x_1|^2+|x_2|^2=1$. Equality in (\ref{eq})
is achieved for $|x_1|=|x_2|=\frac{1}{\sqrt{2}}$ and
$|y_1|=|y_2|=\frac{1}{\sqrt{2}}$ and therefore $\la x\otimes y|{\cal
O}|x\otimes y\ra=2.5$. Similar proof can be carried out for the
lower bound $\gamma_{\rm min}$.
\end{exmp}

\section{Local Numerical Range for $d=2$}

Consider now 2-qubit case corresponding to $d=2$. The set  of
nonlinear equations (\ref{set-y}) reduces to
\begin{eqnarray}
\Big[\frac1{C_{\pm}}w_{11}+C_{\pm}w_{21}-\lambda_2\frac{1+C_{\pm}^2}{C_{\pm}}\Big]y_1
+(u_1+u_2)y_2 &=& 0\,,\\
(u_1+u_2)y_1+\Big[\frac1{C_{\pm}}w_{12}+C_{\pm}w_{22}-\lambda_2\frac{1+C_{\pm}^2}{C_{\pm}}\Big]y_2
&=& 0\ .
\end{eqnarray}
Consider normalized vector
$\ket{q}=(q_1,q_2,q_3,q_4)\in\mathbb{R}^4$. It is separable iff
$q_1q_4=q_2q_3$. Hence, we define
\[
\widetilde{G}(q)=\bra{q}M_{\cal O}'\ket{q}-\lambda_1\Big(\sum_{j=1}^4 q_j^2-1\Big)-
2\lambda_2(q_1q_4-q_2q_3)\,,
\]
where $M_{\cal O}'$ represents matrix of $\mathcal{O}$ in the basis
$\ket{g_i'}\otimes \ket{f_k'}$, that is,
\begin{equation}
    M_{\cal O}'\;=\;\left(\begin{array}{cccc}
w_{11} & 0 & 0 &  u_1 \\
0 & w_{12} & u_2 & 0 \\
0 & u_2 & w_{21} & 0 \\
u_1 & 0 & 0 & w_{22}
\end{array}\right),\qquad u_1,u_2\geq 0\,,\quad w_{ij}\in \mathbb{R}\,.
\end{equation}
Now, $d\widetilde{G}=0$ leads to a linear matrix equation
\begin{equation}\label{M-eq}
{\cal M}\ket{q}=\ket{0}\,,
\end{equation}
where
\[
{\cal M}=\left(
\begin{array}{cccc}
-\lambda_1+w_{11} & 0 & 0 & u_1-\lambda_2 \\
0 & w_{12}-\lambda_1 & u_2+\lambda_2 & 0 \\
0 & u_2+\lambda_2 & -\lambda_1+w_{12} & 0 \\
u_1-\lambda_2 & 0 & 0 & w_{22}-\lambda_1
\end{array}\right).
\]
Obviously, this way we arrive at two separate two-dimensional linear problems.
In order to obtain nonzero solutions the following condition should be
fulfilled:
\[ {\rm det}{\cal M} =
d_1(\lambda_1,\lambda_2)\cdot d_2(\lambda_1,\lambda_2)=0\,,
\]
where
\begin{eqnarray}
d_1(\lambda_1,\lambda_2) &=& u_2^2-w_{12}w_{21}+(w_{12}+w_{21})\lambda_1-\lambda_1^2+2u_2\lambda_2+\lambda_2^2
\,,\\
d_2(\lambda_1,\lambda_2) &=&
u_1^2-w_{11}w_{22}+(w_{11}+w_{22})\lambda_1-\lambda_1^2-2u_1\lambda_2+\lambda_2^2\,.
\end{eqnarray}
Now, assuming
\[
\left\{\begin{array}{l} d_1(\lambda_1,\lambda_2)=0 \\  d_2(\lambda_1,\lambda_2)\neq 0
\end{array}\right.\quad\mbox{or}\quad
\left\{\begin{array}{l} d_1(\lambda_1,\lambda_2)\neq 0 \\  d_2(\lambda_1,\lambda_2)= 0
\end{array}\right.
\]
and using the separability condition,  we get four possible product
vectors $\ket{g_i} \otimes\ket{f_j}$,
\begin{equation}\label{0-1}
\Big\{ {0 \choose 1}\otimes{1 \choose 0},\,{1 \choose 0}\otimes{0
\choose 1},\, {0 \choose 1}\otimes{0 \choose 1},\,{1 \choose
0}\otimes{1 \choose 0} \Big\}\,,
\end{equation}
whereas solving
\begin{equation}\label{condition}
\left\{\begin{array}{l} d_1(\lambda_1,\lambda_2)=0 \\  d_2(\lambda_1,\lambda_2)= 0
\end{array}\right.
\end{equation}
we obtain two solutions $(\lambda_1^+,\lambda_2^+)$ and
$(\lambda_1^-,\lambda_2^-)$ which inserted into (\ref{M-eq}) imply
the following conditions:
\begin{equation}\label{q-set}
\left\{\begin{array}{rcl}
q_1 &=& a_{\pm}q_4\\
q_2 &=& b_{\pm}q_3\\
q_1^2+q_2^2+q_3^2+q_4^2 &=& 1\\
q_1q_4 &=& q_2q_3\,.
\end{array}\right.
\end{equation}
Solving (\ref{q-set}) and factorizing $\ket{q}=\ket{x}\otimes\ket{y}$ we arrive at
\begin{equation}\label{q}
\ket{q}=\left(\!\!\begin{array}{c}
\sqrt{\frac{\xi_{\pm}}{1+\xi_{\pm}}}\\ {\frac{1}{\sqrt{1+\xi_{\pm}}}}\end{array}
\!\right)\otimes \left(\!\!\begin{array}{c}
\sqrt{\frac{\kappa_{\pm}}{1+\kappa_{\pm}}}\\{\frac{1}{\sqrt{1+\kappa_{\pm}}}}\end{array}
\!\right)
\,,
\end{equation}
where
\[ a_{\pm}=\frac{l_{\pm}}{m_{\pm}}\,,\qquad
b_{\pm}=\frac{g_{\pm}}{h_{\pm}}\,,\qquad
\kappa_{\pm}=\frac{a_{\pm}}{b_{\pm}}\,,\qquad \xi_{\pm}=a_{\pm}\cdot b_{\pm}\,,
\]
and
{\scriptsize
\begin{eqnarray*}
l_{\pm} &=& 2 u_1^4+8 u_1^3 u_2+2 u_2^4\pm \left(u_1+u_2\right) \left(-w_{11}+w_{12}+w_{21}-w_{22}\right)\sqrt{\Delta} \\
&& +u_2^2 \left(-w_{11}^2-2 w_{12} w_{21}+w_{11} \left(w_{12}+w_{21}\right)+\left(w_{12}+w_{21}\right) w_{22}-w_{22}^2\right) \\
&& +2 u_1 u_2 \left(4 u_2^2-w_{11}^2-2 w_{12} w_{21}+w_{11} \left(w_{12}+w_{21}\right)+\left(w_{12}+w_{21}\right) w_{22}-w_{22}^2\right)\\
&& +u_1^2 \left(12 u_2^2-w_{11}^2-2 w_{12} w_{21}+w_{11} \left(w_{12}+w_{21}\right)+\left(w_{12}+w_{21}\right) w_{22}-w_{22}^2\right), \\
m_{\pm} &=&\left(u_1+u_2\right) \Big(\pm 2 \left(u_1+u_2\right)\sqrt{\Delta}+\left(w_{11}-w_{12}\right) \left(w_{11}-w_{21}\right) \left(w_{11}-w_{12}-w_{21}+w_{22}\right) \\
&& +u_1^2 \left(-3 w_{11}+w_{12}+w_{21}+w_{22}\right)+2 u_1 u_2 \left(-3 w_{11}+w_{12}+w_{21}+w_{22}\right)+u_2^2 \left(-3 w_{11}+w_{12}+w_{21}+w_{22}\right)\Big)\,, \\
g_{\pm} &=& 2 u_1^4+8 u_1^3 u_2+2 u_2^4\pm\left(u_1+u_2\right)\left(w_{11}-w_{12}-w_{21}+w_{22}\right)\sqrt{\Delta} \\
&&+u_2^2 \left(-w_{12}^2-w_{21}^2+w_{11} \left(w_{12}+w_{21}-2 w_{22}\right)+\left(w_{12}+w_{21}\right) w_{22}\right)\\
&& +2 u_1 u_2 \left(4 u_2^2-w_{12}^2-w_{21}^2+w_{11} \left(w_{12}+w_{21}-2 w_{22}\right)+\left(w_{12}+w_{21}\right) w_{22}\right)\\
&&+u_1^2 \left(12 u_2^2-w_{12}^2-w_{21}^2+w_{11} \left(w_{12}+w_{21}-2 w_{22}\right)+\left(w_{12}+w_{21}\right) w_{22}\right), \\
h_{\pm} &=& \left(u_1+u_2\right) \Big(\pm 2 \left(u_1+u_2\right)\sqrt{ \Delta}+\left(w_{11}-w_{12}\right) \left(w_{12}-w_{22}\right) \left(w_{11}-w_{12}-w_{21}+w_{22}\right)\\
&&+u_1^2 \left(w_{11}-3 w_{12}+w_{21}+w_{22}\right)+2 u_1 u_2
\left(w_{11}-3 w_{12}+w_{21}+w_{22}\right)+u_2^2 \left(w_{11}-3
w_{12}+w_{21}+w_{22}\right)\Big)\,,
\end{eqnarray*}
}%
with {\scriptsize
\begin{equation*}\label{}
\Delta = \left(\left(u_1+u_2\right){}^2+\left(w_{11}-w_{21}\right)
\left(w_{12}-w_{22}\right)\right)
\left(\left(u_1+u_2\right){}^2+\left(w_{11}-w_{12}\right)
\left(w_{21}-w_{22}\right)\right)\ .
\end{equation*}  }
 Note that, in order to have real components of $\ket{q}$,
\begin{equation}\label{condi}
\xi_{\pm}\geq 0\,,\qquad
\kappa_{\pm}\geq 0\,,
\end{equation}
should be fulfilled. As a consequence, either both $a_{\pm}$, $b_{\pm}$ are nonnegative or both are non-positive.
Finally, for $\ket{q}$ given by (\ref{q}) we obtain
\begin{eqnarray}\label{q-value}
\bra{q}M_{\cal O}'\ket{q}\;\equiv\; F_{\pm} &=& \frac{2(u_1+u_2)\sqrt{\xi_{\pm}\kappa_{\pm}}+
\xi_{\pm}\kappa_{\pm}w_{11}+\xi_{\pm}w_{12}+\kappa_{\pm}w_{21}+w_{22}}{
(1+\kappa_{\pm})(1+\xi_{\pm})} \\
&=& \frac{2(u_1+u_2)|a_{\pm}|+
a_{\pm}^2w_{11}+\xi_{\pm}w_{12}+\kappa_{\pm}w_{21}+w_{22}}{
1+\xi_{\pm}+\kappa_{\pm}+a_{\pm}^2}\,.\nonumber
\end{eqnarray}
Taking into account  vectors (\ref{0-1}) one obtains
\begin{equation}\label{e-value}
\bra{g_i\otimes f_j}M_{\cal O}'\ket{g_i \otimes f_j}=w_{ij}\,.
\end{equation}
Hence, LNR of the circulant observable ${\cal O}$ is given by $[\gamma_{\min},
\gamma_{\max}]$, where
\begin{eqnarray}\label{gmax}
\gamma_{\min} &=& \min\Big\{w_{ij},F_{\pm}\Big\} \\
\gamma_{\max} &=& \max\Big\{w_{ij},F_{\pm}\Big\}
\label{gmin}
\end{eqnarray}
To summarize, in order to calculate LNR for a given
$\mathbb{C}^2\otimes \mathbb{C}^2$ circulant operator, we propose
the following procedure
\begin{enumerate}
\item if in a given basis a matrix representation of an operator $M_{\cal O}$ has complex or negative off-diagonal entries
then change the basis due to Proposition 1 and bring the matrix to
the real form,
\item determine real vectors $\ket{x}$ and $\ket{y}$ (see (\ref{q}))  together with
$F_{\pm}$ and compare these values with diagonal elements of
$M_{\cal O}'$. Then LNR$\mathcal{O})=[\gamma_{\min},\gamma_{\max}]$,
where $\gamma_{\min}$ and $\gamma_{\max}$ are defined in
(\ref{gmax}) and (\ref{gmin}), respectively.
\end{enumerate}

\begin{exmp}\rm  As  an illustration let us consider a two-parameter family of
matrices $Q_{t,s}$, $t,s\geq 0$, analyzed in \cite{Z-1},
\[
Q_{t,s}=\left(\begin{array}{cccc}
2 & 0 & 0 & t \\
0 & 1 & s & 0 \\
0 & s & -1 & 0 \\
t & 0 & 0 & -2 \\
\end{array}\right).\]
Denoting by $p=t+s\geq 0$ one obtains
\begin{eqnarray*}
\Delta &=& (1+p^2)(9+p^2) \\
a_{\pm} &=& \frac{4p\pm\sqrt{\Delta}}{p^2-3} \\
b_{\pm} &=& \frac{2p \pm\sqrt{\Delta}}{p^2+3} \\
\kappa_{\pm} &=& \frac{p^4+2p^2+9\pm 2\sqrt{\Delta}}{(p^2-3)(p^2+3)}\\
\xi_{\pm} &=& \frac{p^4+18p^2+9\pm 6\sqrt{\Delta}}{(p^2-3)(p^2+3)}\,.
\end{eqnarray*}
Note that $b_+\geq 0$ and $b_-\leq 0$, hence $a_+\geq 0$ and $a_-\leq 0$.
Finally, $\ket{x}$ and $\ket{y}$ are real under  the condition $p\geq\sqrt{3}$ (see
(\ref{condi})) and using (\ref{q-value}) we arrive at
\[
F_{\pm}=\pm\frac{\sqrt{\Delta}}{2p}\,.
\]
Because the maximal  and minimal values of $w_{ij}$
are equal to 2 and $-2$, respectively, due to (\ref{gmax}) and (\ref{gmin}) we get
\begin{displaymath}
\gamma_{\max} = \left\{ \begin{array}{cl}
2 & \textrm{for $0\leq p < \sqrt{3}$}\\[1ex]
\displaystyle
\frac{1}{2p}\sqrt{\Delta} & \textrm{for $p\geq\sqrt{3}$}
\end{array} \right.
\end{displaymath}
and $\gamma_{\min}=-\gamma_{\max}$
in complete agreement with the result of \cite{Z-1}.
\end{exmp}

\section*{Appendix}

We are going to carry out an extremalization procedure of
(\ref{forma}) with two constraints $|x|=1$, $|y|=1$ using a Lagrange
function $G=F-\lambda_1(|x|^2-1)-\lambda_2(|y|^2-1)$. As a result we
get the following equations:
\begin{eqnarray}
    \pfrac{G}{x_i} &=& x_i\Big[\sum_{k=1}^dw_{ik}y_k^2-\lambda_1\Big]+
    x_{i+1}\sum_{k=1}^dy_ky_{k+1}u_k\cos\varphi_k\;=\;0\,,\quad i=1,2 \label{poly-x}\\
    \pfrac{G}{y_k} &=& y_k\Big[\sum_{i=1}^2w_{ik}x_i^2-\lambda_2\Big]+
    x_1x_2\Big(y_{k-1}u_{k-1}\cos\varphi_{k-1}+y_{k+1}u_k\cos\varphi_k
    \Big)   =0\,,\; k=1,\ldots,d\qquad\mbox{} \label{poly-y}\\
    \pfrac{G}{\beta_1} &=& x_1x_2\sum_{k=1}^dy_ky_{k+1}u_k\sin\varphi_k\;=\;0\,,\\
    \pfrac{G}{\beta_k} &=& x_1x_2\Big(y_ky_{k+1}u_k\sin\varphi_k-y_{k-1}y_{k}u_{k-1}\sin\varphi_{k-1}\Big)\;=\;0\,,\quad
    k=2,\ldots,d\,.
\end{eqnarray}
From the last two equations one obtains in a generic case, i.e.,
when $x_i\neq 0$, and $y_k\neq 0$, the following set of equations
\begin{equation}\label{set}
    \left\{\begin{array}{rcl}
    \displaystyle\sum_{k=1}^dz_k &=&0 \\
    z_{k-1}-z_{k} &=& 0\,,\qquad k=2,\ldots,d\,.
    \end{array}\right.
\end{equation}
with $z_k=y_ky_{k+1}u_k\sin\varphi_k$ or in a matrix notation $\mathbm{W}^T\mathbm{z}=\mathbm{0}$,
where $\mathbm{W}^T$ is a transposition of the matrix given by (\ref{W}).
Now, according to ${\rm det}\mathbm{W}^T=d(-1)^{d+1}\neq 0$, the set of homogeneous equations (\ref{set}) has only zero solution, hence in a generic case, $\sin\varphi_k=0$ for $k=1,\ldots,d$.
The angles $\beta_k$ can now be easily obtained.
It results from $\sin\varphi_k=0$ that
\begin{eqnarray}
     \alpha_1+\beta_1+\beta_2 &=& 0\,,\\ %m_1\pi\,,\\
     \alpha_k+\beta_1-\beta_k+\beta_{k+1} &=& 0\,,\quad k=2,\ldots,d-1 \\
    %m_k\pi\,,\quad k=2,\ldots,d-1 \\
    \alpha_d+\beta_1-\beta_d &=& 0\,.\\ %m_d\pi\,,
\end{eqnarray}
%where $m_j$, $j=1,\ldots,d$ are integers which should be choosen in such a way that the phases
%$\beta_j\in(-\pi,\pi]$.
or in a matrix form
\begin{equation}
    \mathbm{\alpha}=-\mathbm{W}\mathbm{\beta}
\end{equation}
with exactly the same $\mathbm{W}$ as in (\ref{mW}). Hence solutions for $\beta_1,\ldots,\beta_d$ differ
only by a sign from solutions for $\mu_1,\ldots,\mu_d$ (see (\ref{mW})) and
one can easily find that
\begin{eqnarray}
    \beta_1 &=& -\frac{1}{d}\sum_{k=1}^d\alpha_k \;=\;-\mu_1\,,\nonumber \\
    \beta_2 &=& -\alpha_1-\beta_1\;=\;-\mu_2\,,\label{sol-beta}\\
    \beta_{k+1} &=& -\alpha_k-\beta_1+\beta_k\;=\;-\mu_{k+1}\,,\quad  \quad k=2,\ldots,d-1\,.\nonumber
\end{eqnarray}

\section*{Acknowledgments}

This work was partially supported by the Polish
Ministry of Science and Higher Education Grant No 3004/B/H03/2007/33
and Grant UMK 370-F.


\begin{thebibliography}{10}
\bibitem{HJ} R.\,A. Horn, C.\,R. Johnson, {\it Topics in Matrix Analysis},
Cambridge Univ. Press, 1992.
\bibitem{Lax} P.\,D. Lax, {\it Linear Algebra and its Applications}, Wiley, 2007.
\bibitem{Z-1} P. Gawron, Z. Pucha{\l}a, J.\,A. Miszczak, \L. Skowronek, and
K. \.Zyczkowski, {\it Restricted numerical range: a versatile tool in the theory of
quantum information}, arXiv: 0905.3646v2.

\bibitem{Z-2} Z. Pucha{\l}a, P. Gawron,  J.\,A. Miszczak, \L. Skowronek, Man-Duen Choi, and K. \.Zyczkowski, {\it Product numerical range in space with tensor product
structure}, arXiv: 1008.3482v1.

\bibitem{Schulte} T. Schulte-Herbr\"uggen, G. Dirr, U. Helmke, S.\,J. Glaser, \textit{The significance of the $C$-numerical range and the local
$C$-numerical range in quantum control and quantum information},
Linear and Multilinear Algebra {\bf 56},  3 (2008).
\bibitem{Dirr} G. Dirr, U. Helmke, M. Kleinsteuber, T. Schulte-Herbr\"uggen, \textit{Relative $C$-numerical ranges for applications in quantum control and quantum information},
Linear and Multilinear Algebra {\bf 56},  27 (2008).

\bibitem{Toth} O. G\"uhne, G. Toth, {\it Entanglement detection}, Phys. Reports {\bf 474}, 1--75 (2009).
\bibitem{Toth-2} G. Toth, {\it Entanglement witnesses in spin models}, Phys. Rev A
{\bf  71}  010301(R) (2005).
\bibitem{Hor09}
R. Horodecki, P. Horodecki, M. Horodecki, K. Horodecki, {\it Quantum entanglement}, Rev. Mod. Phys.
{\bf 81}, 865--942 (2009).
\bibitem{vogel}
J. Sperling and W. Vogel, \textit{Necessary and sufficient conditions for bipartite entanglement}, Phys. Rev. A {\bf 79}, 022318 (2009).
%\bibitem{peres}
%A. Peres, Phys. Rev. Lett. 77, 1413 (1996).
%\bibitem{Doherty1} A.\,C. Doherty, P.\,A. Parrilo, F.\,M. Spedalieri, {\it A complete family of separability criteria},
%Phys. Rev. A {\bf 69}, 022308 (2004).
%\bibitem{Doherty2}
%A.\,C. Doherty, P.\,A. Parrilo, F.\,M. Spedalieri, {\it Detecting multipartite entanglement},
% Phys. Rev. A, Vol. {\bf 71}, 032333 (2005).
%\bibitem{Lupo}
%P. Aniello, C. Lupo, {\it On the relation between Schmidt coefficients and entanglement},
%Open Sys.  Information Dyn. 16, 127 (2009).
%\bibitem{Hor96}
%M. Horodecki, P. Horodecki, and R. Horodecki, Phys. Lett. A 223, 1 (1996).
\bibitem{darek}
D. Chru\'sci\'nski and A. Kossakowski, {\it Circulant states with positive partial transpose}, Phys. Rev. A 76, 032308 (2007).

\bibitem{Art} D. Chru\'sci\'nski and A. Pittenger, {\it Generalized Circulant Densities and a Sufficient Condition for Separability},
  J. Phys. A: Math. Theor. {\bf 41} (2008) 385301.

\bibitem{Multi} D. Chru\'sci\'nski and A. Kossakowski,
{\it Multipartite Circulant States with Positive Partial Transpose},
Open Sys.  Information Dyn. {\bf 15} (2008) 189-212.

\bibitem{Rau} A. R. P. Rau, \textit{Algebraic characterization of X-states in quantum information},  J. Phys. A: Math. Gen. \textbf{42}, 412002 (2009).
\bibitem{Fan10} F. F. Fanchini, T. Werlang, C. A. Brasil, L. G. E. Arruda, A. O. Caldeira,
\textit{Non-Markovian Dynamics of Quantum Discord}, Phys. Rev. A. \textbf{81}, 052107 (2010).
\bibitem{Maz} M. Ali, A. R. P. Rau, G. Alber, \textit{Quantum discord for two-qubit X-states}, Phys. Rev. A \textbf{81}, 042105 (2010).

\bibitem{Bogna} B. Bylicka, D. Chru\'sci\'nski, {\it Witnessing quantum discord in $2 \times
N$ systems}, Phys. Rev.  A {\bf 81}, 062102 (2010).

\bibitem{YWei} Y.\,S. Weinstein, \textit{Entanglement Sudden Death in Three Qubit X-States}, Phys. Rev. A {\bf 82}, 032326 (2010).

\bibitem{darek2} D. Chru\'sci\'nski, A. Kossakowski, K. M{\l}odawski, and T. Matsuoka,
{\it A class of Bell diagonal states and entanglement witnesses}, Open Sys. Information
Dyn. {\bf 17}, 235 (2010).
%\bibitem{Jur09} J. Jurkowski, D. Chru\'sci\'nski, A. Rutkowski, OSID \textbf{16}, 235 (2009).
\end{thebibliography}
\end{document}